%% file: arxiv.tex
\documentclass{article}

\usepackage{microtype}
\usepackage{graphicx}
\usepackage{subcaption}
\usepackage{booktabs} % for professional tables

\usepackage{hyperref}

\usepackage[preprint]{icml2026}

\usepackage{amsmath}
\usepackage{amssymb}
\usepackage{mathtools}
\usepackage{amsthm}

\usepackage{booktabs}
\usepackage{array}
\usepackage{multirow}
\usepackage{rotating}
\usepackage{xcolor}
\usepackage[capitalize,noabbrev]{cleveref}
\newcommand{\squishlist}{
	\begin{list}{$\bullet$}
		{ \setlength{\itemsep}{1pt}
			\setlength{\parsep}{1pt}
			\setlength{\topsep}{0.5pt}
			\setlength{\partopsep}{0.5pt}
			\setlength{\leftmargin}{1em}
			\setlength{\labelwidth}{1em}
			\setlength{\labelsep}{0.6em}
		}
	}
	\newcommand{\squishend}{
	\end{list}
}

\theoremstyle{plain}

\theoremstyle{definition}

\theoremstyle{remark}

\usepackage[textsize=tiny]{todonotes}

\usepackage{color}
\definecolor{ZpfGreen}{RGB}{0,100,0}
\definecolor{ZpfRed}{RGB}{255,0,102}

\newcommand{\stitle}[1]{\vspace*{0.0em}\noindent{\bf #1.\/}}

\newcommand{\name}{{vLLM-Omni}}
\begin{document}

\twocolumn[
  \icmltitle{\name{}: Fully Disaggregated Serving for Any-to-Any Multimodal Models}



  \icmlsetsymbol{equal}{*}

  \begin{icmlauthorlist}
    \icmlauthor{Peiqi Yin}{equal,hw,cuhk}
    \icmlauthor{Jiangyun Zhu}{equal,cas}
    \icmlauthor{Han Gao}{equal,hw}
    \icmlauthor{Chenguang Zheng}{equal,hw}
    \icmlauthor{Yongxiang Huang}{equal,hw}
    \icmlauthor{Taichang Zhou}{equal,hw}
    \icmlauthor{Ruirui Yang}{equal,hw}
    \icmlauthor{Weizhi Liu}{equal,hw}
    \icmlauthor{Weiqing Chen}{equal,hw}
    \icmlauthor{Canlin Guo}{hw}
    \icmlauthor{Didan Deng}{hw}
    \icmlauthor{Zifeng Mo}{sysu}
    \icmlauthor{Cong Wang}{hw}
    % \icmlauthor{Ziming Huang}{ali}
    % \icmlauthor{Baoyuan Qi}{xiaomi}
    % \icmlauthor{XXX}{hw}
    % \icmlauthor{XXX}{hw}
    % \icmlauthor{...}{hw}
    % \icmlauthor{XXX}{hw}
    \icmlauthor{James Cheng}{cuhk}
    \icmlauthor{Roger Wang}{ind}
    \icmlauthor{Hongsheng Liu}{hw}
  \end{icmlauthorlist}

  \icmlaffiliation{hw}{AI Framework and Data Technology Lab, Huawei}
  \icmlaffiliation{cuhk}{The Chinese University of Hong Kong}
  \icmlaffiliation{cas}{Institute of Software, Chinese Academy of Sciences}
  \icmlaffiliation{sysu}{Sun Yat-sen University}
  \icmlaffiliation{ind}{Independent researcher }
  % \icmlaffiliation{ali}{Alibaba Cloud}
  % \icmlaffiliation{xiaomi}{Xiaomi}

  \icmlcorrespondingauthor{Hongsheng Liu}{liuhongsheng4@huawei.com}

  \icmlkeywords{Machine Learning, ICML}

  \vskip 0.3in
]



\printAffiliationsAndNotice{\textsuperscript{*}Core Contributor. }

\begin{abstract}
Any-to-any multimodal models that jointly handle text, images, video, and audio represent a significant advance in multimodal AI. However, their complex architectures (typically combining multiple autoregressive LLMs, diffusion transformers, and other specialized components) pose substantial challenges for efficient model serving. Existing serving systems are mainly tailored to a single paradigm, such as autoregressive LLMs for text generation or diffusion transformers for visual generation. They lack support for any-to-any pipelines that involve multiple interconnected model components. As a result, developers must manually handle cross-stage interactions, leading to huge performance degradation. 
We present \name{}, a fully disaggregated serving system for any-to-any models. \name{} features a novel stage abstraction that enables users to decompose complex any-to-any architectures into interconnected stages represented as a graph, and a disaggregated stage execution backend that optimizes resource utilization and throughput across stages. Each stage is independently served by an LLM or diffusion engine with per-stage request batching, flexible GPU allocation, and unified inter-stage connectors for data routing. Experimental results demonstrate that \name{} reduces job completion time (JCT) by up to 91.4\% compared to baseline methods. The code is public available at \href{https://github.com/vllm-project/vllm-omni}{https://github.com/vllm-project/vllm-omni}.

\end{abstract}

\input{sections/1_introduction}
\input{sections/2_background}

\input{sections/3_designs}
\input{sections/4_evaluation}
\input{sections/5_related_work}

\input{sections/6_conclusion}

\bibliography{main}
\bibliographystyle{icml2026}

\end{document}

%% file: sections/1_introduction.tex
\section{Introduction}

\begin{figure}[!t]
	\centering
	\includegraphics[width=1\linewidth]{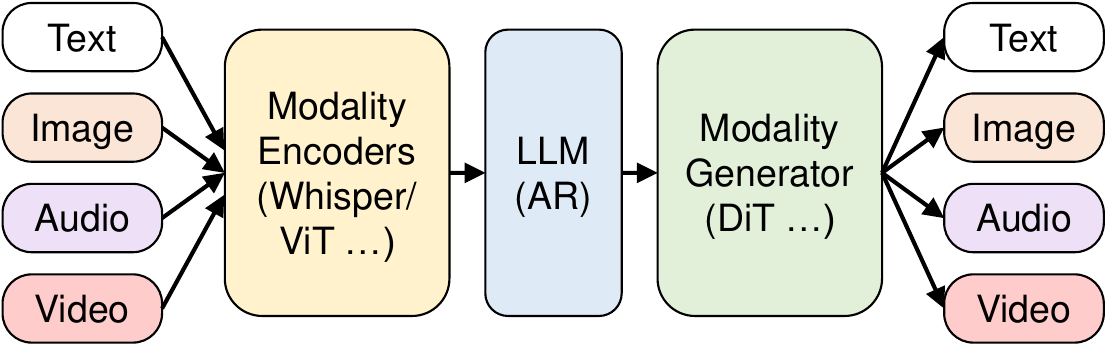}
	\caption{Any-to-any multimodal model architecture.}
	\label{fig:omni overview}
	\vspace{-3mm}
\end{figure}

Traditional large language models (LLMs)~\cite{gpt4, gemini, qwen3} have achieved remarkable performance in language understanding and reasoning tasks, such as question answering~\cite{nqa, longbench}, summarization~\cite{sumbench, qmsum}, and code generation~\cite{code1}. However, LLMs are constrained to text-only modalities, especially for the output. The growing demand for processing multimodal data has motivated the development of multimodal models that extend LLMs with encoders and decoders for diverse modalities~\cite{Qwen-Image, hunyuanimage, GLMImage2026, Ming-Flash-Omni}. A key trend is the emergence of \textit{any-to-any multimodal models}~\cite{Qwen2.5-Omni, Qwen3-Omni, Ming-Flash-Omni, LongCat-Flash-Omni}, unified architectures that seamlessly understand and generate across text, images, video, and audio through end-to-end training (Figure~\ref{fig:omni overview}). This unification enables more flexible cross-modal reasoning and interaction compared to separate understanding and generation pipelines. Recent any-to-any models have achieved SOTA performance across multimodal understanding tasks such as image captioning~\cite{visualgpt, promptcap}, visual question answering~\cite{shao2023prompting, schwenk2022okvqa}, and voice-based assistance~\cite{chen2024internvl, li2024llava}, as well as generation tasks including image editing~\cite{liu2025step1x, kawar2023imagic}, speech translation~\cite{chen2023blaser, chen2024llast}, and text-to-speech synthesis~\cite{voicebench, wang2025spark}.

The emergence of these any-to-any patterns leads to more complex model structures than those of traditional LLMs. For example, to support audio generation, modern any-to-any models such as Qwen-Omni~\cite{Qwen2.5-Omni, Qwen3-Omni} adopt a Thinker–Talker architecture that connects two autoregressive (AR) LLMs, one dedicated to generating text tokens and the other to generating audio tokens. To support image generation, models like GLM-Image~\cite{GLMImage2026} typically use an AR LLM to understand the input and then connect it to a diffusion transformer (DiT)~\cite{DiT} for visual synthesis. More advanced any-to-any models further integrate multiple AR and DiT components to simultaneously support both audio and visual outputs within a unified pipeline~\cite{Ming-Flash-Omni, Ming-Omni}.

Such complexity in model structure introduces substantial challenges for efficient model serving. Existing serving frameworks are typically specialized for a single generation paradigm. LLM serving frameworks such as vLLM~\cite{vllm} and SGLang~\cite{sglang} are optimized for autoregressive LLM decoding and designed for text-only generation, while diffusion serving frameworks~\cite{diffusers, fang2024xdit} are optimized for DiT denoising and designed for image and video generation. As a result, these frameworks lack native support for any-to-any pipelines that involve multiple autoregressive LLMs, DiT models, or other specialized neural components that must interact in customized ways to produce multi-modal outputs. Developers are thus unable to leverage these frameworks and instead resort to custom implementations that are tightly coupled with specific models, resulting in poor performance and limited extensibility.

To address the challenges, we propose \name{}, a fully disaggregated serving system for any-to-any models, featuring a stage abstraction frontend and stage execution backend. Unlike existing LLM serving frameworks that operate on a single AR decoding or DiT denoising stage, \name{} supports complex any-to-any architectures by introducing the concept of a \emph{stage graph}. Through \name{}'s stage abstraction, users can decompose an any-to-any model into multiple stages and explicitly define the pipeline as a stage graph. The nodes represent model stages (e.g., AR or DiT), and edges correspond to user-defined functions that transform and route intermediate data to subsequent stages. 

Given this stage graph specification, \name{} delivers efficient serving for any-to-any models by leveraging a disaggregated stage-execution backend that optimizes execution and resource utilization across stages. 
Each stage is independently served by a specialized execution engine, in which vLLM for LLM stages and a dedicated diffusion engine for DiT stages. Each engine performs per-stage request batching to maximize resource utilization. Users can flexibly allocate computing accelerators and memory resources to each stage according to its computational characteristics.
To support the disaggregated execution across the entire pipeline, \name{} employs a unified connector between stages for flexible and customized intermediate data transfer. 
Our evaluation shows that \name{} achieves substantial performance improvements across diverse multimodal models. For Qwen3-Omni, \name{} reduces job completion time by up to 91.4\% compared to baseline method.

To summarize, we make the following contributions.
\squishlist
\item We introduce \name{}, a fully disaggregated serving system for any-to-any multimodal models. \name{} propose a stage graph abstraction, enabling native support for multi-stage model pipelines.

\item We present a stage execution backend that enables stage-wise optimizations, a unified connector for data transfer, and a diffusion engine for visual generation.

\item Our experiments demonstrate that \name{} consistently outperforms baselines across diverse any-to-any models and tasks.
\squishend

%% file: sections/2_background.tex
\section{Background \& Motivation}

\subsection{Any-to-Any Multimodal Models}
\label{subsec:mllm}

Any-to-any multimodal models~\cite{Qwen3-Omni, Qwen2.5-Omni, Ming-Omni, Ming-Flash-Omni, LongCat-Flash-Omni} extend the capabilities of text-only LLMs by processing and generating outputs across diverse modalities, including text, image, video, and audio. Unlike standard LLMs, Any-to-any models employ specialized architectures for cross-modality understanding and generation. For multimodal comprehension, modern models adopt dedicated encoders (audio encoders such as Whisper~\cite{whisper} and Audio Transformer; vision encoders such as ViT~\cite{Qwen2.5-vl} and SigLIP~\cite{siglip}) that map multimodal inputs into a shared embedding space connected to an LLM backbone. For multimodal generation, the LLM backbone generate embedding outputs and parsed to modality-specific decoders, including text-to-speech models~\cite{ditar, cosyvoice} and image/video generation models~\cite{mmdit, FLUX.1}, to produce diverse output modalities.\footnote{We refer the model with multiple input and output modalities as ``any-to-any'' models in this work. Noted that not all any-to-any models support all input and output modalities. Some models may support multimodal inputs (e.g., text, image, audio) but only generate text and images, or text and audio.} The combination of these components creates increasingly complex architectures that may combine multiple autoregressive (AR), diffusion transformer (DiT), or other specialized generators.

% Figure~\ref{fig:mllm_arch} illustrates the architectural variations across different MLLMs.

\begin{figure}[!t]
	\centering
	\includegraphics[width=1\linewidth]{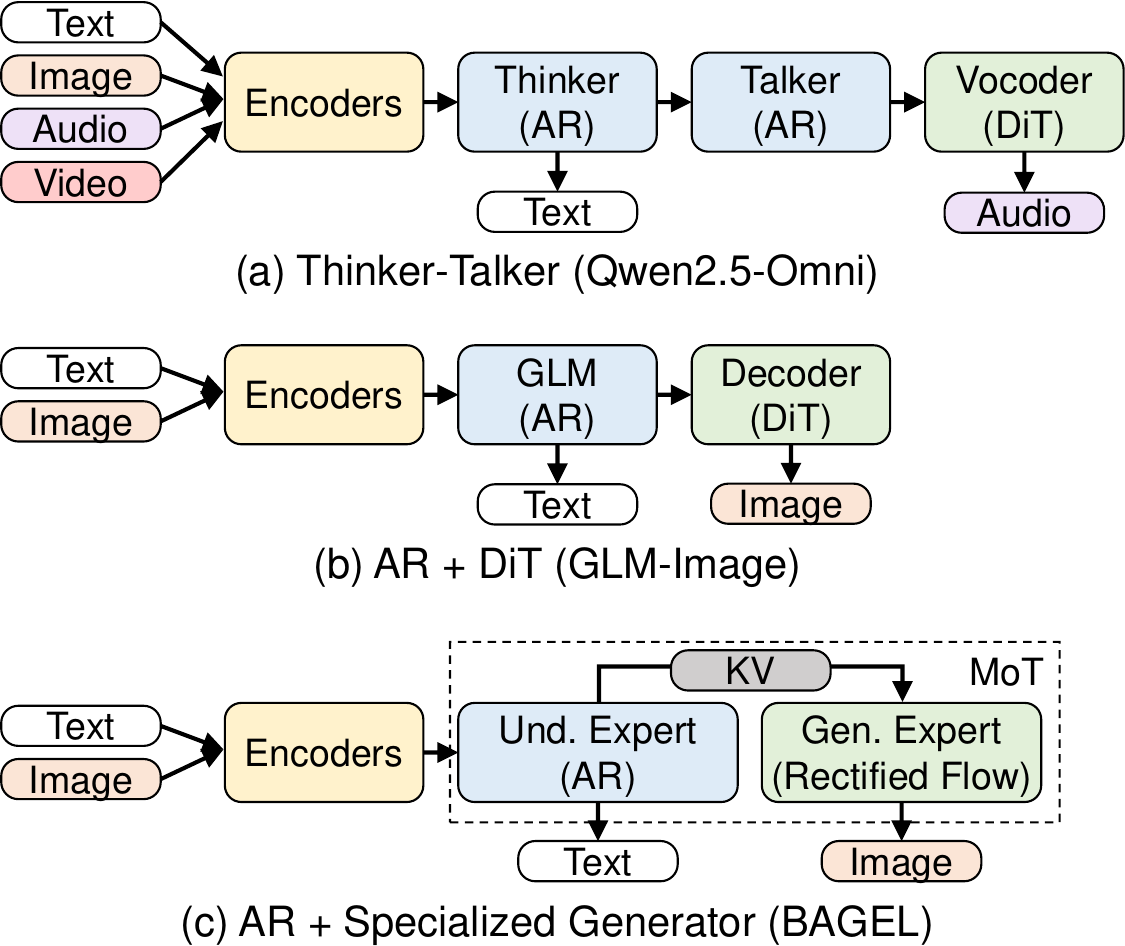}
	\caption{Architecture for existing any-to-any models. (a) Qwen2.5-Omni~\cite{Qwen2.5-Omni}; (b) GLM-Image~\cite{GLMImage2026}; (c) BAGEL~\cite{bagel}.}
	\label{fig:mllm_arch}
	\vspace{-2mm}
\end{figure}

\stitle{Multiple AR LLM decoders}
Some any-to-any models employ multiple autoregressive (AR) LLM decoders within their pipeline. Qwen-Omni series~\cite{Qwen2.5-Omni, Qwen3-Omni} exemplifies this design  (i.e., Figure~\ref{fig:mllm_arch}(a)), supporting inputs across text, image, video, and audio while generating both textual and audio outputs. The model comprises multimodal encoders for input processing, a ``Thinker'' LLM for text generation, a ``Talker'' LLM for audio codec generation, and a ``Vocoder'' module for audio waveform reconstruction.\footnote{For the Vocoder, Qwen2.5-Omni adopts a DiT architecture, while Qwen3-Omni uses a lightweight CNN-based approach.} This Thinker–Talker design incorporates two sequential autoregressive LLM models in the execution pipeline. Similar architectures have been adopted by other any-to-any models, e.g., Ming-Omni series~\cite{Ming-Flash-Omni, Ming-Omni}.

% For omni-modal systems such as Qwen2.5-Omni and Qwen3-Omni, the architectures support inputs across text, image, video, and audio, while generating both textual and auditory outputs. These models typically comprise multimodal encoders for input comprehension, an LLM ``Thinker'' for text generation, an LLM ``Talker'' for audio codec synthesis, and a Code2Wav module to reconstruct the final audio waveform. Specifically, Qwen2.5-Omni employs dense LLMs for both the Thinker and Talker components, utilizing a Diffusion Transformer (DiT) for the Code2Wav module. Conversely, Qwen3-Omni upgrades the Thinker and Talker to Mixture-of-Experts (MoE) architectures and adopts a lightweight convolutional network (ConvNet) for the Code2Wav module.

% Many any-to-any models that support image, video, or audio generation adopt a modular pipeline that separates (i) semantic planning / token generation and (ii) modality-specific synthesis. A common instantiation combines an autoregressive (AR) LLM with a dedicated generator (e.g., diffusion transformers (DiT)~\cite{DiT} or other specialized decoders) to translate high-level semantics into high-fidelity outputs.

\stitle{AR with specialized generators}
Many any-to-any models adopt a modular pipeline that separates (i) semantic generation from (ii) modality-specific synthesis. A common instantiation combines an AR LLM with diffusion transformers (DiT)~\cite{DiT} to translate high-level semantics into high-fidelity outputs. GLM-Image~\cite{GLMImage2026} follows such a hybrid design (Figure~\ref{fig:mllm_arch}(b)): it first employs semantic-VQ~\cite{geng2025x} with a VAE-based encoder~\cite{van2017neural} to extract visual features, and then uses a 9B autoregressive LLM (GLM-4~\cite{glm2024chatglm}) for semantic understanding and token generation. The generated tokens are subsequently consumed by a 7B single-stream DiT decoder to synthesize the final image.

This ``AR + specialized generator'' principle also appears in other any-to-any models with visual or audio outputs~\cite{LongCat-Flash-Omni, Step-audio, ge2024seed, dong2023dreamllm, tong2025metamorph, MiMo-Audio, bagel}. For example, LongCat-Flash-Omni~\cite{LongCat-Flash-Omni} uses a 560B-parameter MoE LLM backbone for autoregressive token generation, followed by a lightweight LSTM/CNN-based audio decoder that reconstructs waveforms in real time. Step-Audio~\cite{Step-audio} employs a 130B-parameter LLM for generate speech token, followed by a hybrid decoder with DiT flow-matching for Mel-spectrograms and neural vocoding for waveforms. BAGEL~\cite{bagel} can likewise be interpreted through this modular lens: its Mixture-of-Transformers (MoT) design separates multimodal semantic understanding from visual generation via different experts, which can be viewed as two stages within a unified model (Figure~\ref{fig:mllm_arch}(c)).

\subsection{Challenges to Existing LLM Serving Frameworks}

Existing open-source LLM serving frameworks, such as vLLM~\cite{vllm}, SGLang~\cite{sglang}, are designed around a \textit{step-centric} paradigm optimized for text-only LLM inference. These frameworks encapsulate iteration logic and attention key-value (KV) cache management within their runtime, enabling model developers to implement only a single forward pass through the \textsf{forward} function. This abstraction is specifically tailored for sequential text generation, where a model produces tokens iteratively from a fixed input prompt to a text output.

However, the emergence of any-to-any multimodal LLMs introduces architectural challenges that expose the limitations of this step-centric design. Any-to-any models typically include multiple model components of different types, such as autoregressive (AR) LLMs, diffusion transformers (DiTs), and other neural network architectures, connected in complex multi-stage pipelines. The step-centric abstraction becomes a fundamental mismatch: it is designed to represent a single forward pass of text generation, not the coordinated execution and data flow across multiple heterogeneous stages. Consequently, existing frameworks like vLLM and SGLang cannot support multimodal generation, as their abstraction cannot express a multi-stage pipeline.
%lack the flexibility to handle data processing between model stages.

Qwen-Omni models adopt the Thinker-Talker architecture with three model components, and their pipeline logic cannot be expressed within the step-centric frameworks. Developers need to first implement the step-centric forward pass for each stage independently, then manually orchestrate the inter-stage transfer outside the serving framework. The workflow proceeds as follows: multimodal inputs are passed to the LLM Thinker stage via the end-to-end \textsf{generate()} function, which executes the encodes and the AR decoding loop to generate output text. Upon completion, the output hidden states are extracted and transformed into input embeddings for the Talker stage. The Talker then executes its own AR generation loop via the customized \textsf{generate()} function. Finally, upon completion of the Talker stage, outputs are passed to the Vocoder stage for waveform reconstruction.

\begin{figure*}[!t]
    \setlength{\abovecaptionskip}{0.1cm}
	\centering
	\includegraphics[width=2.1\columnwidth]{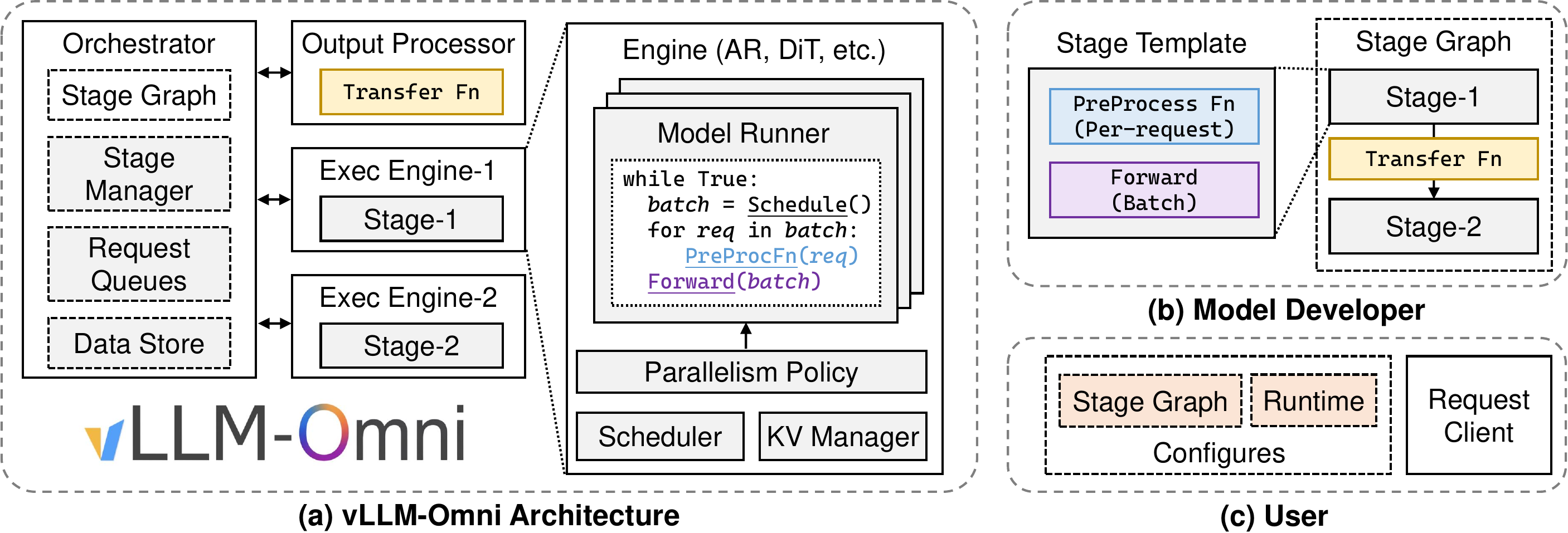}
	\caption{\name{} architecture.}
	\label{fig:omni arch}
	\vspace{-3mm}
\end{figure*}

Such manual implementations incur significant performance penalties. First, serving multimodal generation cannot leverage the efficiency optimizations provided by well-engineered serving frameworks. Existing serving systems are designed with fixed input and output types, making it difficult to deploy on even one particular model stage. Such that, performance optimization techniques such as continuous batching~\cite{yu2022orca} for decoding and chunked prefill~\cite{Sarathi} processing cannot be applied. Second, as model components are implemented and executed together as a monolithic program, computing resources cannot be efficiently allocated across stages, and the overall pipeline cannot be decomposed or dynamically adjusted. This co-location of stages prevents fine-grained resource distribution, further degrading serving performance.

%% file: sections/3_designs.tex
\section{\name{} Designs}

\name{} is a disaggregated serving system for any-to-any multimodal models that enables efficient, scalable inference across heterogeneous model components. In this section, we introduce the designs of our \name{} system: \S~\ref{subsec:overview} provides an architectural overview of the \name{} system; \S~\ref{subsec:abstraction} describes the stage abstraction interface for any-to-any model programming; \S~\ref{subsec:execution} explains the stage execution pipeline and diffusion model integration; \S~\ref{subsec:transfer} presents the data transfer mechanism for stage disaggregation; and \S~\ref{subsec:hardware} discusses the hardware support.
%Section~\ref{subsec:diffusion} discusses optimizations for diffusion model integration;

% We propose \name{}, a general and efficient serving system for omni-modal models, featuring a stage abstraction frontend and stage execution backend. Unlike existing LLM serving frameworks that operate on a single AR decoding or DiT denoising stage, \name{} supports complex MLLM architectures by introducing the concept of a \emph{stage graph}. In \name{}, users decompose an MLLM into multiple stages and explicitly define the pipeline as a stage graph, where nodes represent model stages (e.g., AR or DiT), and edges correspond to user-defined functions that transform and route intermediate data to subsequent stages. Given this stage graph specification, \name{} delivers efficient serving for omni-modal models by leveraging a disaggregated stage-execution backend that optimizes execution and resource utilization across stages.

\subsection{Overview}
\label{subsec:overview}

Figure~\ref{fig:omni arch} illustrates an overview of \name{}.
We show the backend architecture of \name{} in Figure~\ref{fig:omni arch}(a). On the backend, an orchestrator manages the execution of stages and schedules incoming requests. Each stage is served by an independent execution engine, enabling independent stage scaling, resource allocation, and intra-stage request batching. During inference, the model runner iteratively takes batched requests, applies the corresponding \textit{preprocess} function on each scheduled request, and executes one batched forward step for each iteration. A unified connector then transfers intermediate data between stages, enabling fully disaggregated execution across the entire pipeline.

Figure~\ref{fig:omni arch}(b) presents the stage abstraction exposed to model developers: each component of an any-to-any model (e.g., LLM and DiT) is implemented as an independent stage equipped with a customized \textit{preprocess} function and a batched \textit{forward} function. The \textit{preprocess} function enables developers to modify stage inputs with additional data produced by preceding stages. From the endpoint users’ perspective (Figure~\ref{fig:omni arch}(c)), \name{} exposes runtime configurations, including parallelism strategies and memory budgets for different stages, allowing users to tune performance and resource usage without changing model code.

\subsection{Stage Abstraction}
\label{subsec:abstraction}

\name{} provides a flexible and easy-to-use frontend interface for any-to-any model programming, as illustrated by the template in Figure~\ref{fig:omni arch}(b). In this system, users define any-to-any models as a \textit{stage graph}, where nodes represent model stages and edges represent stage-transfer functions. This allows for the decomposition of complex architectures (i.e., autoregressive LLMs, DiTs, or CNN modules) into distinct stages.\footnote{Multimodal encoders can be treated either as a separate stage or as part of the LLM stage.}
Specifically, for each AR stage, users implement a \textit{preprocess} function to modify stage inputs and a \textit{forward} function in the same step-centric manner as in existing LLM serving systems, enabling batched execution within the stage. To manage data flow between these stages, users define stage-transfer functions that control how query states and intermediate data are transformed during transitions. By combining these stage execution and transfer definitions, the stage graph encapsulates the full execution pipeline of the any-to-any model.

\begin{figure}[!t]
	\setlength{\abovecaptionskip}{0.1cm}
	\centering
	\includegraphics[width=1\linewidth]{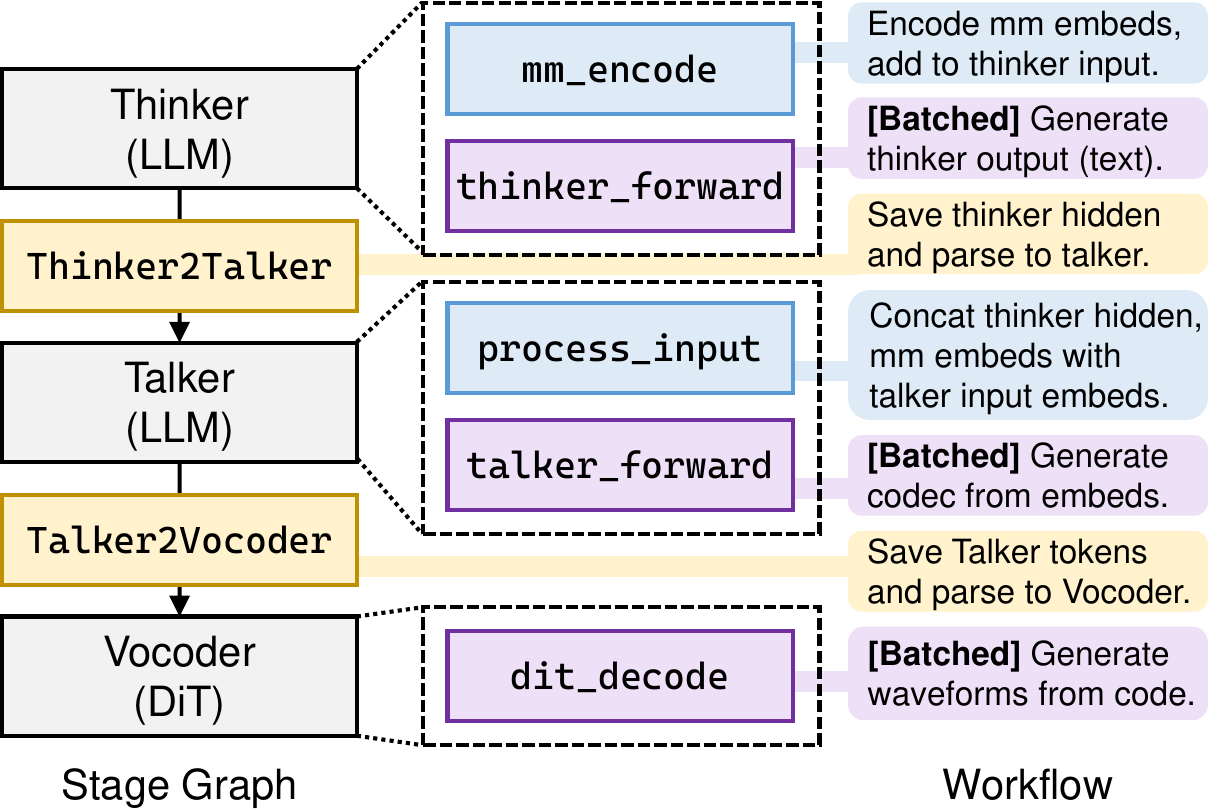}
	\caption{An example implementation of Qwen2.5-Omni (i.e., stage graph and workflow for the model). Qwen3-Omni is similar.}
	\label{fig:omni model}
	\vspace{-2mm}
\end{figure}

We show an example of implementing the Qwen2.5-Omni model using \name{} in Figure~\ref{fig:omni model}. As we discussed in Section~\ref{subsec:mllm}, Qwen2.5-Omni includes three stages: (i) an LLM Thinker for text generation, (ii) an LLM Talker for audio code generation, and (iii) a DiT Vocoder for waveform synthesis. The model takes multimodal inputs, where dedicated encoders transform audio, image, and video into embeddings that are concatenated with the textual inputs. These inputs are first fed into the Thinker LLM to produce textual outputs and corresponding hidden states. Next, the Talker stage receives the Thinker outputs and concatenates the Thinker hidden states and multimodal embeddings with the Talker input embeddings. The Talker LLM then autoregressively generates codec tokens, repeatedly concatenating the Thinker hidden states at each decoding step. Finally, the generated codec sequence is passed to the Vocoder, which produces audio waveforms via DiT denoising.

Under the stage paradigm of \name{}, users implement three types of functions: (i) the \textit{forward} function for each model stage (e.g., \texttt{thinker\_forward}, \texttt{talker\_forward}, \texttt{dit\_decode}); (ii) \textit{preprocess} functions that construct stage inputs (e.g., \texttt{mm\_encode} to obtain multimodal embeddings and concatenate them with Thinker inputs\footnote{In this example, we regard the multimodal encoder as a part of the Thinker stage, follow the implementation of vLLM.}; and \texttt{process\_input} to concatenate Thinker hidden states with Talker input embeddings, and is called in each decode iteration); and (iii) stage-transfer functions between stages (e.g., \texttt{Thinker2Talker} and \texttt{Talker2Vocoder}, they are only called once). In typical use, users define the \textit{forward} and \textit{preprocess} logic for each node, construct a stage graph, and assign transfer functions to the edges. In this way, \name{} decouples any-to-any models into modular stages while still fully exploiting the performance optimizations of underlying serving engines, allowing users to benefit from efficient resource utilization without manually handling batching or scheduling logic.
%\pq{TODO: discuss other models?}

\subsection{Stage Execution}
\label{subsec:execution}

Given a stage graph and user-specified runtime configurations, \name{} first initializes a set of execution engines, where each engine hosts a single model component, loads the corresponding model parameters, and starts serving according to the configured parallelism policy and memory budget. An orchestrator process is then launched to manage request routing and data exchange across stages.

As each stage runs on an independent engine, \name{} can naturally disaggregate execution across different stages from the complex any-to-any model structures. Engines can be configured with different parameters and accelerator resources according to the characteristics and demands of each underlying model stage. For example, in the three-stage Qwen3-Omni pipeline, the Thinker model is the largest (30B), so more accelerator memory can be allocated to the Thinker stage; the Talker model is smaller but more compute-intensive, so it can be assigned less memory but higher parallelism and more accelerators. Each engine can also enable standard serving optimizations such as chunked prefill~\cite{Sarathi} and runtime execution-graph compilation, inheriting the performance benefits of LLM serving systems.

\stitle{AR stage support}
We use the vLLM~\cite{vllm} engine for serving AR stages.
For each engine, batching scheduling, KV-cache management, and model execution are handled independently by its own scheduler, KV manager, and model runner. The model runner implements the logic for executing the customized preprocess function for each request, which enables flexible composition of multi-stage models. Specifically, we introduce a predefined dictionary for storing intermediate per-request data that users can access and update on both the transform functions and the preprocess functions. The preprocess function is invoked at every iteration, since some stages need to combine outputs from previous stages with the current forward inputs at each decoding step (e.g., Talker stage for Qwen-Omni). The output processor is responsible for applying the transform function, storing the resulting data in CPU memory, and then transferring it to the device that hosts the next stage.

\stitle{DiT stage support}
\name{} integrates a dedicated diffusion engine seamlessly into its multi-stage pipeline architecture.\footnote{Could implement other generation stages within this engine.} By treating the diffusion process as a distinct node within the stage graph, the system extends its core disaggregated serving principles to diffusion workflows, ensuring efficient inference for audio, image, and video generation tasks. To maximize throughput and reduce latency, the engine implements a suite of optimization techniques, including advanced attention mechanisms (flash attention~\cite{dao2022flashattention}, SAGE attention~\cite{zhang2024sageattention}, and TurboAttention~\cite{kang2025turboattention}), caching strategies for the iterative denoising process (TeaCache~\cite{teacache}, cache-dit~\cite{cache-dit}), and parallelization approaches such as RingAttention-based context parallelism~\cite{ring} and Ulysses sequence parallelism. These optimizations enable \name{} to serve diffusion models with improved throughput and reduced latency compared to baseline implementations.

Ultimately, the diffusion engine empowers \name{} to support a wide range of state-of-the-art diffusion models, including text-to-image generation (Z-Image~\cite{Z-Image}, Qwen-Image~\cite{Qwen-Image}, Flux~\cite{FLUX.1}), image editing tools (Qwen-Image-Edit~\cite{Qwen-Image}, LongCat-Image-Edit~\cite{LongCat-Image}), and video generation variants (Wan2.2 variants~\cite{wan}, HunyuanVideo~\cite{Hunyuanvideo}).

% \pq{TODO: discuss the orchestrator?}
% The orchestrator manage the stage for each request. 

\textbf{Streaming stage output.}
In multi-stage pipeline execution, some downstream stages do not require complete outputs from their preceding stages to begin computation. In the Qwen-Omni pipeline, for example, the Vocoder can start processing audio generation as soon as the Talker produces initial tokens, rather than waiting for the entire sequence to be completed. To support this pattern, \name{} implements streaming stage output, where intermediate results are transferred to downstream stages incrementally as they become available. The output processor asynchronously streams partial outputs, such as newly generated tokens or embeddings, to the next stage while the upstream stage continues execution. By enabling overlapped execution across stages, streaming stage output reduces time-to-first-token (TTFT) for the final output and supports streaming responses to users without requiring stages to be tightly synchronized.

\subsection{Disaggregated Data Transfer}
\label{subsec:transfer}

\begin{figure}[!t]
	\centering
	\includegraphics[width=1\linewidth]{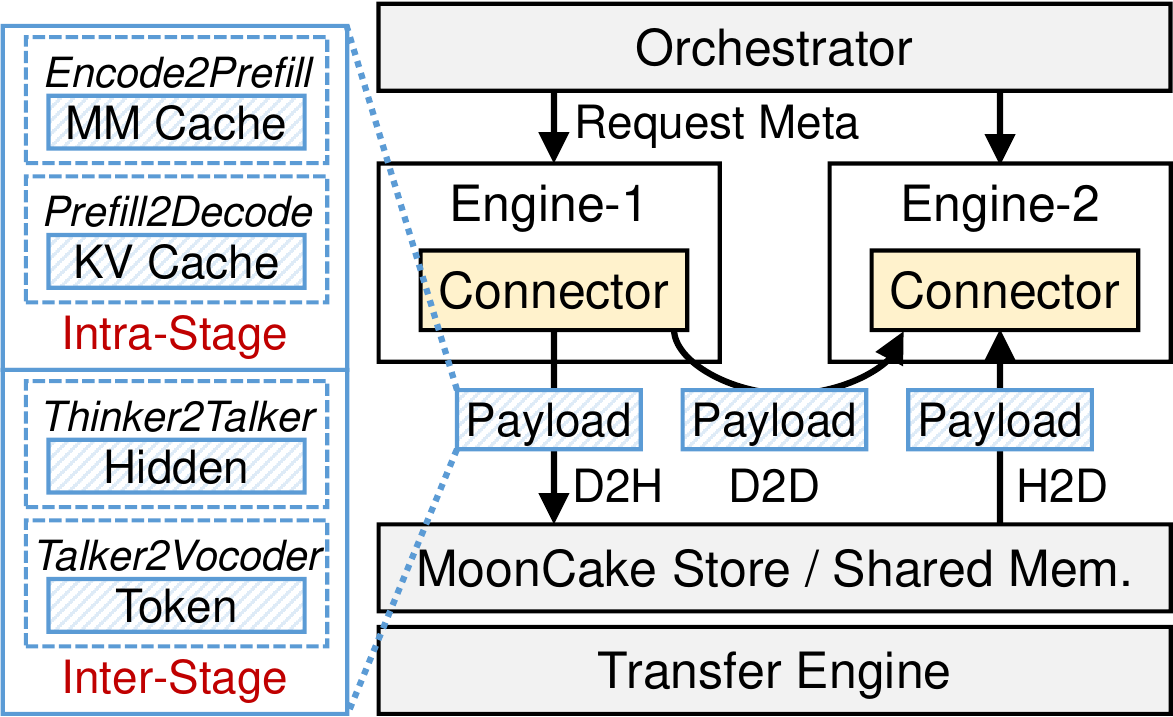}
	\caption{Disaggregated data transfer with unified connector.}
	\label{fig:omni connector}
	\vspace{-2mm}
\end{figure}

\name{} supports disaggregated data transfer through a \textit{unified connector} interface that decouples transport from model logic. Inspired by vLLM’s KV cache transfer mechanisms for prefill–decode disaggregation, \name{} generalizes the connector interface to handle a broader range of data objects, including embeddings, hidden states, and audio or image tensors. This unified connector layer is responsible for data movement between stages, enabling full disaggregation of components such as encoders, prefill, decode, and modality-specific generators. 

The unified connector is responsible for transferring data between model stages. For single-node deployments, it provides low-latency transfer by using inline control queues for small payloads and system shared memory for larger ones. In distributed multi-node settings, we leverage Ray~\cite{ray} to orchestrate cross-node execution. A Mooncake-based connector~\cite{mooncake} complements this by enabling TCP- or RDMA-based transport, allowing stages on different servers to exchange data via a common \textsf{put}/\textsf{get} interface while passing only lightweight metadata through the control plane. By separating stage execution from data transport and allowing per-edge connector setting, \name{} flexibly supports heterogeneous deployment topologies and scales any-to-any pipelines across nodes without changing the programming model.

The unified connector also handles intra-stage transfers, including KV cache between prefill and decode and multimodal (MM) cache between encoder and prefill. This design remains compatible with existing EPD (encode–prefill–decode) disaggregation~\cite{epd}.

% \subsection{Diffusion Support}
% \label{subsec:diffusion}

\subsection{Hardware Support}
\label{subsec:hardware}

\name{} supports diverse hardware platforms to enable flexible any-to-any model serving. Built upon vLLM's hardware plugin architecture, \name{} achieves cross-platform compatibility through a decoupled plugin mechanism that allows for registering hardware-specific implementations independently.

%% file: sections/4_evaluation.tex
\section{Experimental Evaluation}

\subsection{Experiment Settings}

% \stitle{Datasets}

\stitle{Models}
\name{} extends the serving capabilities of vLLM to any-to-any tasks with support for multimodal outputs. Specifically, we evaluate the system using a set of representative and state-of-the-art models: (i) Thinker-Talker architecture that connects two autoregressive (AR) models within the execution pipeline, i.e., Qwen3-Omni~\cite{Qwen3-Omni} and Qwen2.5-Omni~\cite{Qwen2.5-Omni}, which are any-to-any models with both text and audio outputs. (ii) two-stage models that couple an AR LLM with additional modality-specific components for generation: e.g., BAGEL~\cite{bagel} employs a Mixture-of-Transformer-Experts design with separate experts for multimodal understanding and generation (paired with separate visual encoders), while MiMo-Audio~\cite{MiMo-Audio} combines a patch encoder, an AR LLM backbone, and a patch decoder to generate audio tokens autoregressively. (iii) diffusion models with image or video outputs, i.e., Qwen-Image~\cite{Qwen-Image}, Qwen-Image-Edit~\cite{Qwen-Image}, and Wan2.2 series~\cite{wan}, built primarily on diffusion transformers.

\stitle{Baseline Systems}
For Qwen-Omni models, we use their default Hugging Face Transformers~\cite{Transformers} implementations to evaluate offline inference performance, as vLLM and SGLang only support their thinker part. For BAGEL and MiMo-Audio, we adopt its original implementation as our baseline. For Diffusion-based models, we adopt Diffusers library~\cite{diffusers} as our baseline.

\stitle{Metrics}
For models with audio outputs (i.e., the Qwen-Omni series and MiMo-Audio), we primarily evaluate Real-Time Factor (RTF) and Job Completion Time (JCT) as our performance metrics. RTF is defined as the ratio between the end-to-end processing time and the duration of the generated audio. JCT measures the end-to-end latency of each request, from submission to completion.
For the Qwen-Omni models, we further report Tokens Per Second (TPS) for both the thinker and the talker components. Thinker TPS denotes the throughput of generated text tokens per second, while talker TPS denotes the throughput of generated audio tokens per second. For visual generation models, we adopt JCT as our main performance metric.

\stitle{Testbed}
The experiments were conducted on a server equipped with two accelerator devices (80GB memory each), 24 CPU cores, and 192 GB of system memory. The environment was configured with a virtual setup, and vLLM version 0.12.0 was used.

\subsection{End-to-End Performance}

\begin{figure}[!t]
	\setlength{\abovecaptionskip}{0.06cm}
	\centering
	\includegraphics[width=1\linewidth]{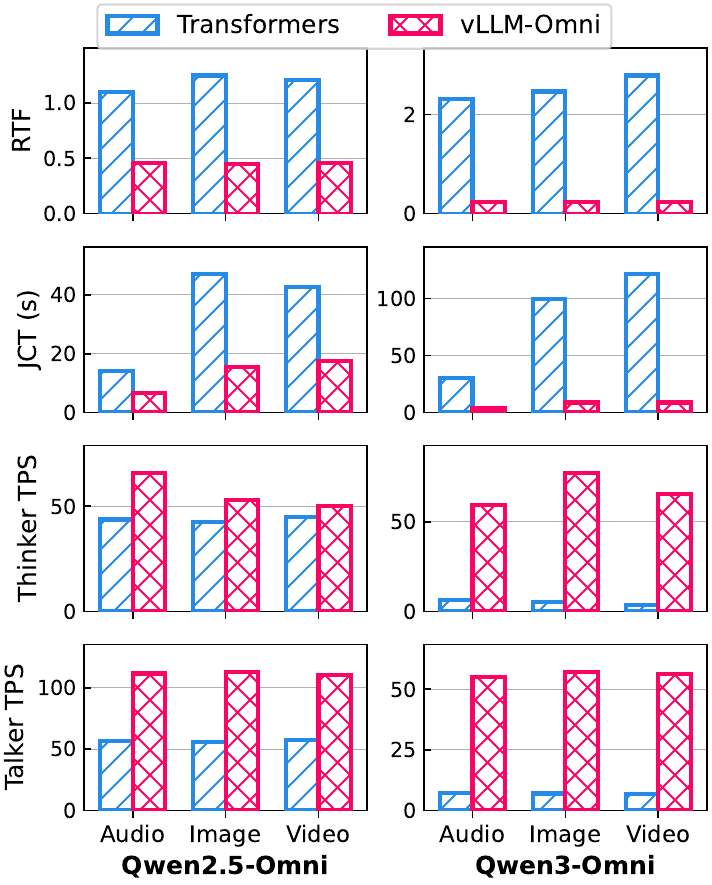}
	\caption{End-to-end results on Qwen-Omni models.}
	\label{fig:omni result}
	\vspace{-2mm}
\end{figure}

\stitle{Thinker-Talker architecture}
Figure~\ref{fig:omni result} shows the end-to-end performance of \name{} and the baseline on the Qwen-Omni series. We use the \textit{librispeech\_asr}~\cite{panayotov2015librispeech}, \textit{food101}~\cite{bossard2014food}, and \textit{ucf101-subset}~\cite{soomro2012ucf101} datasets as the audio, image, and video inputs, respectively. All evaluations use the first 100 queries from each dataset as input. The experiments are run on 2 accelerators with 80GB memory, where the baseline uses the default tensor-parallel configuration of the Transformers implementation. \name{} applies tensor parallelism to the Thinker across both accelerators, while placing the Talker on device-1 and the Vocoder on device-0.

For Qwen2.5-Omni, compared with the baseline Transformers implementation, \name{} reduces RTF by 61.4\% and JCT by 61.6\%. For Qwen3-Omni, \name{} reduces RTF by 90.7\% and JCT by 91.4\%. These results indicate that \name{} delivers substantial end-to-end performance gains over the existing implementation.
To further analyze the bottlenecks, we report Thinker TPS and Talker TPS. \name{} achieves 1.29× and 1.97× higher TPS on the Thinker and Talker, respectively, for Qwen2.5-Omni, and 12.97× and 7.98× higher Thinker TPS and Talker TPS for Qwen3-Omni. These results show that \name{} significantly accelerates Qwen3-Omni, with more than 10× speedup on the Thinker.
This large improvement on Qwen3-Omni is attributed to the additional optimizations implemented in \name{}, while the baseline implementation does not fully exploit modern LLM serving techniques such as execution graph compilation. Since Qwen3-Omni has a much larger Thinker model (30B) than Qwen2.5-Omni (7B), \name{} can better amortize its optimized execution pipeline and thus achieves higher relative gains.

\begin{figure}[!t]
	\setlength{\abovecaptionskip}{0.06cm}
	\centering
	\includegraphics[width=1\linewidth]{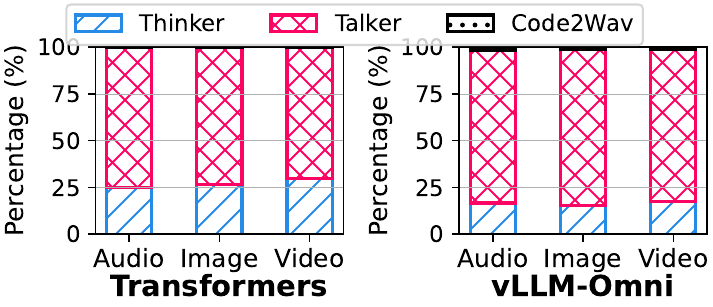}
	\caption{Execution time decompose for Qwen3-Omni.}
	\label{fig:time decompose}
	\vspace{-2mm}
\end{figure}

Figure~\ref{fig:time decompose} illustrates the time decomposition across different stages for the Qwen3-Omni model. The results show that, for both systems, the Talker stage accounts for most of the overall latency because it needs to generate substantially more tokens for audio than the Thinker does for text.
For example, on video-input tasks, the average input token count (including video tokens) is 841.6, the average number of output text tokens is 150.9, while the average number of output audio tokens reaches 545.4. Consequently, the Talker runs many more decoding iterations than the Thinker and results in longer latency.

% \stitle{Online Serving Performance}
% \pq{TODO: run it}

\stitle{BAGEL Model}
We evaluated BAGEL on an accelerator with 80 GB memory on Vbench~\cite{huang2024vbench}. For image generation tasks with 1024×1024 resolution, the baseline implementation achieves a JCT (Job Completion Time) of 23.12s for text-to-image (T2I) and 41.39s for image-to-image (I2I). Our method reduces JCT to 9.64s for T2I and 11.12s for I2I, yielding a 2.40× speedup for T2I and 3.72× speedup for I2I over the baseline.

\stitle{Mimo-Audio model} We evaluated Mimo-Audio on one accelerator on SeedTTS (text-to-speech)~\cite{seedtts}. The baseline implementation achieves an RTF of 1.39, while our method reduces RTF to 0.60 without execution-graph compilation and to 0.12 with graph compilation, yielding an 11.58× speedup over the baseline.

\subsection{Micro Experiments}

\stitle{Diffusion engine}
We compare \name{} with Diffusers on DiT-based image and video generation models using the VBench~\cite{huang2024vbench} dataset. For text-to-image and image-to-image generation, we employ Qwen-Image and Qwen-Image-Edit, respectively. For video generation, we use Wan2.2-T2V and Wan2.2-I2V for text-to-video and image-to-video tasks. The output image resolution is 1024x1024 for Qwen-Image and 480x640 for Wan2.2, with 80 output frames. Results demonstrate that \name{} consistently outperforms Diffusers with a 1.26x overall speedup. This performance gain stems from \name{}'s diffusion engine, which reuses operator optimizations and the flash-attention backend from vLLM, enabling efficient execution across diverse generation tasks.

\begin{figure}[!t]
	\setlength{\abovecaptionskip}{0.06cm}
	\centering
	\includegraphics[width=1\linewidth]{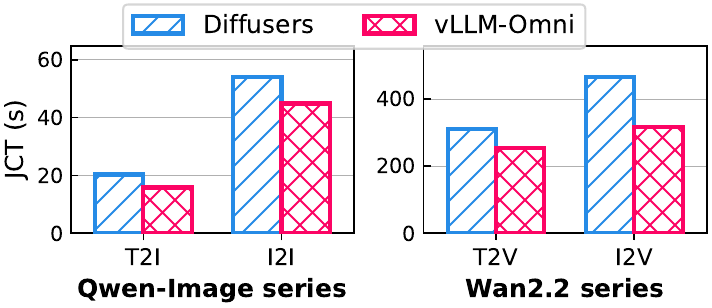}
	\caption{End-to-end results on DiT-based models.}
	\label{fig:diffusion}
	\vspace{-2mm}
\end{figure}

\begin{table}[t]
	\setlength{\abovecaptionskip}{0.05cm}
  \centering
  \caption{Data transfer time with using \name{}'s unified connector. The model is Qwen2.5-Omni.}
  \label{tab:unified-connector-overhead}
  \renewcommand{\arraystretch}{0.9} % 默认 1，减小行距[web:1]
  \begin{tabular}{l|c|c}
    \toprule
    Latency (ms) & Thinker2Talker & Talker2Vocoder \\
    \midrule
    Shared Memory & 5.49 & 0.53 \\
    Mooncake & 8.28 & 3.34 \\
    \bottomrule
  \end{tabular}
	\vspace{-3mm}
\end{table}

\stitle{Unified connector}
We evaluate the data transfer overhead of \name{}'s unified connector in Table~\ref{tab:unified-connector-overhead}. Results demonstrate that the connector overhead is negligible relative to overall inference latency (tens of seconds), making it a practical solution for disaggregated execution. Despite the minimal performance cost, the unified connector provides substantial flexibility by abstracting data movement across heterogeneous deployment topologies.

%% file: sections/5_related_work.tex
\section{Related Work}

\stitle{LLM serving systems}
Existing systems for autoregressive LLM serving, such as vLLM~\cite{vllm} and SGLang~\cite{sglang}, provide efficient support for text-only or multimodal input LLMs with text-only output. These systems offer a step-wise frontend interface that abstracts away serving optimizations from users. The execution backend integrates LLM execution optimizations, including attention implementations (e.g., paged attention~\cite{vllm}, flash attention~\cite{dao2022flashattention}), KV cache management, and prefix tree caching~\cite{sglang}. They incorporate various optimizations to achieve lower latency and higher throughput, such as chunked prefill~\cite{Sarathi}, continuous batching~\cite{yu2022orca}, and Prefill-Decode (PD)~\cite{zhong2024distserve} disaggregation. Additionally, these systems support multiple parallelism strategies, such as data parallelism and tensor parallelism. Since \name{}'s execution engine extends vLLM's engine, it inherits these optimizations for AR stages within any-to-any pipelines.

\stitle{Multimodal model serving}
Current LLM serving systems~\cite{vllm, sglang, Transformers} support LLM with multimodal inputs~\cite{Qwen2.5-vl, chen2024internvl, li2024llava}, with optimizations like Encode-Prefill-Decode (EPD) disaggregation~\cite{epd, qiu2025modserve} and multimodal embedding cache~\cite{wan2024look}. They still focus on text-only output scenarios and lack the support for models with multimodal output. However, they focus exclusively on text-only output and lack support for multimodal generation. Separately, diffusion-based systems~\cite{fang2024xdit, diffusers, huang2022prodiff, huang2025ddit} efficiently accelerate visual and audio generation through optimizations like quantized attention~\cite{kang2025turboattention, zhang2024sageattention}, parallel denoising~\cite{ring}, and caching strategies~\cite{cache-dit, teacache, agarwal2024approximate}. Yet these frameworks are specialized for diffusion models and struggle with complex architectures, particularly when integrating heavy LLM-based text encoders. In contrast, \name{} provides unified support for complex pipelines by seamlessly combining autoregressive models with diffusion models, enabling efficient serving for any-to-any multimodal models. Such ability is essential for deploying next-generation any-to-any models at scale.

%% file: sections/6_conclusion.tex
\section{Conclusion}

In this paper, we presented \name{}, a serving system for efficient deployment of any-to-any multimodal models. The key insight of our work is decomposing complex any-to-any model architectures into a stage graph, where each stage can be independently optimized and executed. Through our disaggregated stage execution backend, \name{} enables efficient serving support for diverse any-to-any models. Our experimental results demonstrate substantial improvements compared to existing approaches.

%% file: main.bib
@article{nqa,
  title={Natural questions: a benchmark for question answering research},
  author={Kwiatkowski, Tom and Palomaki, Jennimaria and Redfield, Olivia and Collins, Michael and Parikh, Ankur and Alberti, Chris and Epstein, Danielle and Polosukhin, Illia and Devlin, Jacob and Lee, Kenton and others},
  journal={Transactions of the Association for Computational Linguistics},
  volume={7},
  pages={453--466},
  year={2019},
  publisher={MIT Press One Rogers Street, Cambridge, MA 02142-1209, USA journals-info~…}
}

@inproceedings{longbench,
  title={Longbench: A bilingual, multitask benchmark for long context understanding},
  author={Bai, Yushi and Lv, Xin and Zhang, Jiajie and Lyu, Hongchang and Tang, Jiankai and Huang, Zhidian and Du, Zhengxiao and Liu, Xiao and Zeng, Aohan and Hou, Lei and others},
  booktitle={Proceedings of the 62nd annual meeting of the association for computational linguistics (volume 1: Long papers)},
  pages={3119--3137},
  year={2024}
}

@article{sumbench,
  title={Benchmarking large language models for news summarization},
  author={Zhang, Tianyi and Ladhak, Faisal and Durmus, Esin and Liang, Percy and McKeown, Kathleen and Hashimoto, Tatsunori B},
  journal={Transactions of the Association for Computational Linguistics},
  volume={12},
  pages={39--57},
  year={2024},
  publisher={MIT Press One Broadway, 12th Floor, Cambridge, Massachusetts 02142, USA~…}
}

@inproceedings{qmsum,
  title={QMSum: A new benchmark for query-based multi-domain meeting summarization},
  author={Zhong, Ming and Yin, Da and Yu, Tao and Zaidi, Ahmad and Mutuma, Mutethia and Jha, Rahul and Hassan, Ahmed and Celikyilmaz, Asli and Liu, Yang and Qiu, Xipeng and others},
  booktitle={Proceedings of the 2021 Conference of the North American Chapter of the Association for Computational Linguistics: Human Language Technologies},
  pages={5905--5921},
  year={2021}
}

@article{code1,
  title={Llm-based test-driven interactive code generation: User study and empirical evaluation},
  author={Fakhoury, Sarah and Naik, Aaditya and Sakkas, Georgios and Chakraborty, Saikat and Lahiri, Shuvendu K},
  journal={IEEE Transactions on Software Engineering},
  year={2024},
  publisher={IEEE}
}

@article{gpt4,
  title={Gpt-4 technical report},
  author={Achiam, Josh and Adler, Steven and Agarwal, Sandhini and Ahmad, Lama and Akkaya, Ilge and Aleman, Florencia Leoni and Almeida, Diogo and Altenschmidt, Janko and Altman, Sam and Anadkat, Shyamal and others},
  journal={arXiv preprint arXiv:2303.08774},
  year={2023}
}

@article{gemini,
  title={Gemini: a family of highly capable multimodal models},
  author={Anil, Rohan and Borgeaud, Sebastian and Alayrac, Jean-Baptiste and Yu, Jiahui and Soricut, Radu and Schalkwyk, Johan and Dai, Andrew M and Hauth, Anja and Millican, Katie and others},
  journal={arXiv preprint arXiv:2312.11805},
  year={2023}
}

@article{qwen3,
  title={Qwen3 technical report},
  author={Yang, An and Li, Anfeng and Yang, Baosong and Zhang, Beichen and Hui, Binyuan and Zheng, Bo and Yu, Bowen and Gao, Chang and Huang, Chengen and Lv, Chenxu and others},
  journal={arXiv preprint arXiv:2505.09388},
  year={2025}
}

@article{voicebench,
  title={Voicebench: Benchmarking llm-based voice assistants},
  author={Chen, Yiming and Yue, Xianghu and Zhang, Chen and Gao, Xiaoxue and Tan, Robby T and Li, Haizhou},
  journal={arXiv preprint arXiv:2410.17196},
  year={2024}
}

@article{visualgpt,
  title={Visual chatgpt: Talking, drawing and editing with visual foundation models},
  author={Wu, Chenfei and Yin, Shengming and Qi, Weizhen and Wang, Xiaodong and Tang, Zecheng and Duan, Nan},
  journal={arXiv preprint arXiv:2303.04671},
  year={2023}
}

@article{promptcap,
  title={Promptcap: Prompt-guided task-aware image captioning},
  author={Hu, Yushi and Hua, Hang and Yang, Zhengyuan and Shi, Weijia and Smith, Noah A and Luo, Jiebo},
  journal={arXiv preprint arXiv:2211.09699},
  year={2022}
}

@inproceedings{shao2023prompting,
  title={Prompting large language models with answer heuristics for knowledge-based visual question answering},
  author={Shao, Zhenwei and Yu, Zhou and Wang, Meng and Yu, Jun},
  booktitle={Proceedings of the IEEE/CVF Conference on computer vision and pattern recognition},
  pages={14974--14983},
  year={2023}
}

@inproceedings{schwenk2022okvqa,
  title={A-okvqa: A benchmark for visual question answering using world knowledge},
  author={Schwenk, Dustin and Khandelwal, Apoorv and Clark, Christopher and Marino, Kenneth and Mottaghi, Roozbeh},
  booktitle={European conference on computer vision},
  pages={146--162},
  year={2022},
  organization={Springer}
}

@inproceedings{chen2024internvl,
  title={Internvl: Scaling up vision foundation models and aligning for generic visual-linguistic tasks},
  author={Chen, Zhe and Wu, Jiannan and Wang, Wenhai and Su, Weijie and Chen, Guo and Xing, Sen and Zhong, Muyan and Zhang, Qinglong and Zhu, Xizhou and Lu, Lewei and others},
  booktitle={Proceedings of the IEEE/CVF conference on computer vision and pattern recognition},
  pages={24185--24198},
  year={2024}
}

@article{li2024llava,
  title={LLaVA-OneVision: Easy Visual Task Transfer},
  author={Li, Bo and Zhang, Yuanhan and Guo, Dong and Zhang, Renrui and Li, Feng and Zhang, Hao and Zhang, Kaichen and Zhang, Peiyuan and Li, Yanwei and Liu, Ziwei and others},
  journal={Transactions on Machine Learning Research}
}

@article{liu2025step1x,
  title={Step1x-edit: A practical framework for general image editing},
  author={Liu, Shiyu and Han, Yucheng and Xing, Peng and Yin, Fukun and Wang, Rui and Cheng, Wei and Liao, Jiaqi and Wang, Yingming and Fu, Honghao and Han, Chunrui and others},
  journal={arXiv preprint arXiv:2504.17761},
  year={2025}
}

@inproceedings{kawar2023imagic,
  title={Imagic: Text-based real image editing with diffusion models},
  author={Kawar, Bahjat and Zada, Shiran and Lang, Oran and Tov, Omer and Chang, Huiwen and Dekel, Tali and Mosseri, Inbar and Irani, Michal},
  booktitle={Proceedings of the IEEE/CVF conference on computer vision and pattern recognition},
  pages={6007--6017},
  year={2023}
}

@inproceedings{chen2023blaser,
  title={BLASER: A text-free speech-to-speech translation evaluation metric},
  author={Chen, Mingda and Duquenne, Paul-Ambroise and Andrews, Pierre and Kao, Justine and Mourachko, Alexandre and Schwenk, Holger and Costa-juss{\`a}, Marta R},
  booktitle={Proceedings of the 61st Annual Meeting of the Association for Computational Linguistics (Volume 1: Long Papers)},
  pages={9064--9079},
  year={2023}
}

@inproceedings{chen2024llast,
  title={LLaST: Improved End-to-end Speech Translation System Leveraged by Large Language Models},
  author={Chen, Xi and Zhang, Songyang and Bai, Qibing and Chen, Kai and Nakamura, Satoshi},
  booktitle={Findings of the Association for Computational Linguistics ACL 2024},
  pages={6976--6987},
  year={2024}
}

@article{wang2025spark,
  title={Spark-tts: An efficient llm-based text-to-speech model with single-stream decoupled speech tokens},
  author={Wang, Xinsheng and Jiang, Mingqi and Ma, Ziyang and Zhang, Ziyu and Liu, Songxiang and Li, Linqin and Liang, Zheng and Zheng, Qixi and Wang, Rui and Feng, Xiaoqin and others},
  journal={arXiv preprint arXiv:2503.01710},
  year={2025}
}

@article{Qwen2.5-Omni,
  title={Qwen2.5-omni technical report},
  author={Xu, Jin and Guo, Zhifang and He, Jinzheng and Hu, Hangrui and He, Ting and Bai, Shuai and Chen, Keqin and Wang, Jialin and Fan, Yang and Dang, Kai and others},
  journal={arXiv preprint arXiv:2503.20215},
  year={2025}
}

@article{Qwen3-Omni,
  title={Qwen3-Omni Technical Report},
  author={Jin Xu and Zhifang Guo and Hangrui Hu and Yunfei Chu and Xiong Wang and Jinzheng He and Yuxuan Wang and Xian Shi and Ting He and Xinfa Zhu and Yuanjun Lv and Yongqi Wang and Dake Guo and He Wang and Linhan Ma and Pei Zhang and Xinyu Zhang and Hongkun Hao and Zishan Guo and Baosong Yang and Bin Zhang and Ziyang Ma and Xipin Wei and Shuai Bai and Keqin Chen and Xuejing Liu and Peng Wang and Mingkun Yang and Dayiheng Liu and Xingzhang Ren and Bo Zheng and Rui Men and Fan Zhou and Bowen Yu and Jianxin Yang and Le Yu and Jingren Zhou and Junyang Lin},
  journal={arXiv preprint arXiv:2509.17765},
  year={2025}
}

@article{hunyuanimage,
  title={HunyuanImage 3.0 Technical Report},
  author={Cao, Siyu and Chen, Hangting and Chen, Peng and Cheng, Yiji and Cui, Yutao and Deng, Xinchi and Dong, Ying and Gong, Kipper and Gu, Tianpeng and Gu, Xiusen and others},
  journal={arXiv preprint arXiv:2509.23951},
  year={2025}
}

@misc{Qwen-Image,
      title={Qwen-Image Technical Report}, 
      author={Chenfei Wu and Jiahao Li and Jingren Zhou and Junyang Lin and Kaiyuan Gao and Kun Yan and Sheng-ming Yin and Shuai Bai and Xiao Xu and Yilei Chen and Yuxiang Chen and Zecheng Tang and Zekai Zhang and Zhengyi Wang and An Yang and Bowen Yu and Chen Cheng and Dayiheng Liu and Deqing Li and Hang Zhang and Hao Meng and Hu Wei and Jingyuan Ni and Kai Chen and Kuan Cao and Liang Peng and Lin Qu and Minggang Wu and Peng Wang and Shuting Yu and Tingkun Wen and Wensen Feng and Xiaoxiao Xu and Yi Wang and Yichang Zhang and Yongqiang Zhu and Yujia Wu and Yuxuan Cai and Zenan Liu},
      year={2025},
      eprint={2508.02324},
      archivePrefix={arXiv},
      primaryClass={cs.CV},
      url={https://arxiv.org/abs/2508.02324}, 
}

@article{Ming-Flash-Omni,
  title={Ming-Flash-Omni: A Sparse, Unified Architecture for Multimodal Perception and Generation},
  author={Ma, Bowen and Zou, Cheng and Yan, Canxiang and Jin, Chunxiang and Shen, Chunjie and Lian, Chenyu and Zheng, Dandan and Wang, Fudong and Xu, Furong and others},
  journal={arXiv preprint arXiv:2510.24821},
  year={2025}
}

@article{Ming-Omni,
  title={Ming-Omni: A Unified Multimodal Model for Perception and Generation},
  author={Gong, Biao and Zou, Cheng and Zheng, Chuanyang and Zhou, Chunluan and Yan, Canxiang and Jin, Chunxiang and Shen, Chunjie and Zheng, Dandan and Wang, Fudong and others},
  journal={arXiv preprint arXiv:2506.09344},
  year={2025}
}

@inproceedings{vllm,
  title={Efficient memory management for large language model serving with pagedattention},
  author={Kwon, Woosuk and Li, Zhuohan and Zhuang, Siyuan and Sheng, Ying and Zheng, Lianmin and Yu, Cody Hao and Gonzalez, Joseph and Zhang, Hao and Stoica, Ion},
  booktitle={Proceedings of the 29th symposium on operating systems principles},
  pages={611--626},
  year={2023}
}

@article{sglang,
  title={Sglang: Efficient execution of structured language model programs},
  author={Zheng, Lianmin and Yin, Liangsheng and Xie, Zhiqiang and Sun, Chuyue Livia and Huang, Jeff and Yu, Cody Hao and Cao, Shiyi and Kozyrakis, Christos and Stoica, Ion and Gonzalez, Joseph E and others},
  journal={Advances in neural information processing systems},
  volume={37},
  pages={62557--62583},
  year={2024}
}

@inproceedings{DiT,
  title={Scalable diffusion models with transformers},
  author={Peebles, William and Xie, Saining},
  booktitle={Proceedings of the IEEE/CVF international conference on computer vision},
  pages={4195--4205},
  year={2023}
}

@article{LongCat-Flash-Omni,
  title={LongCat-Flash-Omni Technical Report},
  author={Wang, Bairui and Xiao, Bin and Zhang, Bo and Rong, Bolin and Chen, Borun and Wan, Chang and Zhang, Chao and Huang, Chen and Chen, Chen and others},
  journal={arXiv preprint arXiv:2511.00279},
  year={2025}
}

@inproceedings{whisper,
  title={Robust speech recognition via large-scale weak supervision},
  author={Radford, Alec and Kim, Jong Wook and Xu, Tao and Brockman, Greg and McLeavey, Christine and Sutskever, Ilya},
  booktitle={International conference on machine learning},
  pages={28492--28518},
  year={2023},
  organization={PMLR}
}

@article{Qwen2.5-vl,
  title={Qwen2. 5-vl technical report},
  author={Bai, Shuai and Chen, Keqin and Liu, Xuejing and Wang, Jialin and Ge, Wenbin and Song, Sibo and Dang, Kai and Wang, Peng and Wang, Shijie and Tang, Jun and others},
  journal={arXiv preprint arXiv:2502.13923},
  year={2025}
}

@article{siglip,
  title={Siglip 2: Multilingual vision-language encoders with improved semantic understanding, localization, and dense features},
  author={Tschannen, Michael and Gritsenko, Alexey and Wang, Xiao and Naeem, Muhammad Ferjad and Alabdulmohsin, Ibrahim and Parthasarathy, Nikhil and Evans, Talfan and Beyer, Lucas and Xia, Ye and Mustafa, Basil and others},
  journal={arXiv preprint arXiv:2502.14786},
  year={2025}
}

@inproceedings{ditar,
  title={DiTAR: Diffusion Transformer Autoregressive Modeling for Speech Generation},
  author={Jia, Dongya and Chen, Zhuo and Chen, Jiawei and Du, Chenpeng and Wu, Jian and Cong, Jian and Zhuang, Xiaobin and Li, Chumin and Wei, Zhen and Wang, Yuping and others},
  booktitle={Forty-second International Conference on Machine Learning}
}

@article{cosyvoice,
  title={Cosyvoice: A scalable multilingual zero-shot text-to-speech synthesizer based on supervised semantic tokens},
  author={Du, Zhihao and Chen, Qian and Zhang, Shiliang and Hu, Kai and Lu, Heng and Yang, Yexin and Hu, Hangrui and Zheng, Siqi and Gu, Yue and Ma, Ziyang and others},
  journal={arXiv preprint arXiv:2407.05407},
  year={2024}
}

@article{FLUX.1,
  title={FLUX. 1 Kontext: Flow Matching for In-Context Image Generation and Editing in Latent Space},
  author={Labs, Black Forest and Batifol, Stephen and Blattmann, Andreas and Boesel, Frederic and Consul, Saksham and Diagne, Cyril and Dockhorn, Tim and English, Jack and English, Zion and Esser, Patrick and others},
  journal={arXiv preprint arXiv:2506.15742},
  year={2025}
}

@article{bagel,
  title   = {Emerging Properties in Unified Multimodal Pretraining},
  author  = {Deng, Chaorui and Zhu, Deyao and Li, Kunchang and Gou, Chenhui and Li, Feng and Wang, Zeyu and Zhong, Shu and Yu, Weihao and Nie, Xiaonan and Song, Ziang and Shi, Guang and Fan, Haoqi},
  journal = {arXiv preprint arXiv:2505.14683},
  year    = {2025}
}

@article{wan,
  title={Wan: Open and advanced large-scale video generative models},
  author={Wang, Ang and Ai, Baole and Wen, Bin and Mao, Chaojie and Xie, Chen-Wei and Chen, Di and Yu, Feiwu and Zhao, Haiming and Yang, Jianxiao and others},
  journal={arXiv preprint arXiv:2503.20314},
  year={2025}
}

@inproceedings{mmdit,
  title={Scaling rectified flow transformers for high-resolution image synthesis},
  author={Esser, Patrick and Kulal, Sumith and Blattmann, Andreas and Entezari, Rahim and M{\"u}ller, Jonas and Saini, Harry and Levi, Yam and Lorenz, Dominik and Sauer, Axel and Boesel, Frederic and others},
  booktitle={Forty-first international conference on machine learning},
  year={2024}
}

@article{MiMo-Audio,
  title={MiMo-Audio: Audio Language Models are Few-Shot Learners},
  author={Zhang, Dong and Wang, Gang and Xue, Jinlong and Fang, Kai and Zhao, Liang and Ma, Rui and Ren, Shuhuai and Liu, Shuo and Guo, Tao and Zhuang, Weiji and others},
  journal={arXiv preprint arXiv:2512.23808},
  year={2025}
}

@inproceedings{Transformers,
  title={Transformers: State-of-the-art natural language processing},
  author={Wolf, Thomas and Debut, Lysandre and Sanh, Victor and Chaumond, Julien and Delangue, Clement and Moi, Anthony and Cistac, Pierric and Rault, Tim and Louf, Remi and Funtowicz, Morgan and others},
  booktitle={Proceedings of the 2020 conference on empirical methods in natural language processing: system demonstrations},
  pages={38--45},
  year={2020}
}

@inproceedings{mooncake,
  title={Mooncake: Trading more storage for less computation—a $\{$KVCache-centric$\}$ architecture for serving $\{$LLM$\}$ chatbot},
  author={Qin, Ruoyu and Li, Zheming and He, Weiran and Cui, Jialei and Ren, Feng and Zhang, Mingxing and Wu, Yongwei and Zheng, Weimin and Xu, Xinran},
  booktitle={23rd USENIX Conference on File and Storage Technologies (FAST 25)},
  pages={155--170},
  year={2025}
}

@inproceedings{epd,
  title={Efficiently Serving Large Multimodal Models Using EPD Disaggregation},
  author={Singh, Gursimran and Wang, Xinglu and Hu, Yifan and Yu, Timothy Tin Long and Xing, Linzi and Jiang, Wei and Wang, Zhefeng and Xiaolong, Bai and Li, Yi and Xiong, Ying and others},
  booktitle={Forty-second International Conference on Machine Learning}
}

@article{Sarathi,
  title={Sarathi: Efficient llm inference by piggybacking decodes with chunked prefills},
  author={Agrawal, Amey and Panwar, Ashish and Mohan, Jayashree and Kwatra, Nipun and Gulavani, Bhargav S and Ramjee, Ramachandran},
  journal={arXiv preprint arXiv:2308.16369},
  year={2023}
}

@article{LongCat-Image,
      title={LongCat-Image Technical Report},
      author={Hanghang Ma and Haoxian Tan and Jiale Huang and Junqiang Wu and Jun-Yan He and Lishuai Gao and Songlin Xiao and Xiaoming Wei and Xiaoqi Ma and Xunliang Cai and Yayong Guan and Jie Hu},
	    journal={arXiv preprint arXiv:2512.07584},
      year={2025}
}

@article{Z-Image,
  title={Z-Image: An Efficient Image Generation Foundation Model with Single-Stream Diffusion Transformer},
  author={Cai, Huanqia and Cao, Sihan and Du, Ruoyi and Gao, Peng and Hoi, Steven and Huang, Shijie and Hou, Zhaohui and Jiang, Dengyang and Jin, Xin and Li, Liangchen and others},
  journal={arXiv preprint arXiv:2511.22699},
  year={2025}
}

@article{Hunyuanvideo,
  title={Hunyuanvideo: A systematic framework for large video generative models},
  author={Kong, Weijie and Tian, Qi and Zhang, Zijian and Min, Rox and Dai, Zuozhuo and Zhou, Jin and Xiong, Jiangfeng and Li, Xin and Wu, Bo and Zhang, Jianwei and others},
  journal={arXiv preprint arXiv:2412.03603},
  year={2024}
}

@inproceedings{dao2022flashattention,
  title={Flash{A}ttention: Fast and Memory-Efficient Exact Attention with {IO}-Awareness},
  author={Dao, Tri and Fu, Daniel Y. and Ermon, Stefano and Rudra, Atri and R{\'e}, Christopher},
  booktitle={Advances in Neural Information Processing Systems (NeurIPS)},
  year={2022}
}

@inproceedings{zhang2024sageattention,
  title={SageAttention: Accurate 8-Bit Attention for Plug-and-play Inference Acceleration},
  author={Zhang, Jintao and Zhang, Pengle and Zhu, Jun and Chen, Jianfei and others},
  booktitle={The Thirteenth International Conference on Learning Representations}
}

@article{kang2025turboattention,
  title={TurboAttention: Efficient attention approximation for high throughputs llm},
  author={Kang, Hao and Bharadwaj, Srikant and Hensman, James and Krishna, Tushar and R{\"u}hle, Victor and Rajmohan, Saravan},
  journal={Proceedings of Machine Learning and Systems},
  volume={7},
  year={2025}
}

@article{seedtts,
  title={Seed-tts: A family of high-quality versatile speech generation models},
  author={Anastassiou, Philip and Chen, Jiawei and Chen, Jitong and Chen, Yuanzhe and Chen, Zhuo and Chen, Ziyi and Cong, Jian and Deng, Lelai and Ding, Chuang and Gao, Lu and others},
  journal={arXiv preprint arXiv:2406.02430},
  year={2024}
}

@inproceedings{teacache,
  title={Timestep Embedding Tells: It's Time to Cache for Video Diffusion Model},
  author={Liu, Feng and Zhang, Shiwei and Wang, Xiaofeng and Wei, Yujie and Qiu, Haonan and Zhao, Yuzhong and Zhang, Yingya and Ye, Qixiang and Wan, Fang},
  booktitle={Proceedings of the Computer Vision and Pattern Recognition Conference},
  pages={7353--7363},
  year={2025}
}

@misc{cache-dit,
  title={cache-dit: A PyTorch-native and Flexible Inference Engine with Hybrid Cache Acceleration and Parallelism for DiTs.},
  url={https://github.com/vipshop/cache-dit.git},
  note={Open-source software available at https://github.com/vipshop/cache-dit.git},
  author={DefTruth, vipshop.com},
  year={2025}
}

@inproceedings{Ring,
  title={RingAttention with Blockwise Transformers for Near-Infinite Context},
  author={Liu, Hao and Zaharia, Matei and Abbeel, Pieter},
  booktitle={The Twelfth International Conference on Learning Representations}
}

@inproceedings{zhong2024distserve,
  title={$\{$DistServe$\}$: Disaggregating prefill and decoding for goodput-optimized large language model serving},
  author={Zhong, Yinmin and Liu, Shengyu and Chen, Junda and Hu, Jianbo and Zhu, Yibo and Liu, Xuanzhe and Jin, Xin and Zhang, Hao},
  booktitle={18th USENIX Symposium on Operating Systems Design and Implementation (OSDI 24)},
  pages={193--210},
  year={2024}
}

@inproceedings{yu2022orca,
  title={Orca: A distributed serving system for $\{$Transformer-Based$\}$ generative models},
  author={Yu, Gyeong-In and Jeong, Joo Seong and Kim, Geon-Woo and Kim, Soojeong and Chun, Byung-Gon},
  booktitle={16th USENIX Symposium on Operating Systems Design and Implementation (OSDI 22)},
  pages={521--538},
  year={2022}
}

@inproceedings{qiu2025modserve,
  title={Modserve: Modality-and stage-aware resource disaggregation for scalable multimodal model serving},
  author={Qiu, Haoran and Biswas, Anish and Zhao, Zihan and Mohan, Jayashree and Khare, Alind and Choukse, Esha and Goiri, {\'I}{\~n}igo and Zhang, Zeyu and Shen, Haiying and Bansal, Chetan and others},
  booktitle={Proceedings of the 2025 ACM Symposium on Cloud Computing},
  pages={817--830},
  year={2025}
}

@inproceedings{wan2024look,
  title={LOOK-M: Look-Once Optimization in KV Cache for Efficient Multimodal Long-Context Inference},
  author={Wan, Zhongwei and Wu, Ziang and Liu, Che and Huang, Jinfa and Zhu, Zhihong and Jin, Peng and Wang, Longyue and Yuan, Li},
  booktitle={Findings of the Association for Computational Linguistics: EMNLP 2024},
  pages={4065--4078},
  year={2024}
}

@inproceedings{agarwal2024approximate,
  title={Approximate caching for efficiently serving $\{$Text-to-Image$\}$ diffusion models},
  author={Agarwal, Shubham and Mitra, Subrata and Chakraborty, Sarthak and Karanam, Srikrishna and Mukherjee, Koyel and Saini, Shiv Kumar},
  booktitle={21st USENIX Symposium on Networked Systems Design and Implementation (NSDI 24)},
  pages={1173--1189},
  year={2024}
}

@article{fang2024xdit,
  title={xDiT: an Inference Engine for Diffusion Transformers (DiTs) with Massive Parallelism},
  author={Fang, Jiarui and Pan, Jinzhe and Sun, Xibo and Li, Aoyu and Wang, Jiannan},
  journal={arXiv preprint arXiv:2411.01738},
  year={2024}
}

@misc{diffusers,
  author = {Patrick von Platen and Suraj Patil and Anton Lozhkov and Pedro Cuenca and Nathan Lambert and Kashif Rasul and Mishig Davaadorj and Dhruv Nair and Sayak Paul and William Berman and Yiyi Xu and Steven Liu and Thomas Wolf},
  title = {Diffusers: State-of-the-art diffusion models},
  year = {2022},
  publisher = {GitHub},
  journal = {GitHub repository},
  howpublished = {\url{https://github.com/huggingface/diffusers}}
}

@online{GLMImage2026,
  title={GLM-Image: Auto-regressive for Dense-knowledge and High-fidelity Image Generation},
  author={Z.AI},
  year={2026},
  url={https://z.ai/blog/glm-image}
}

@article{glm2024chatglm,
  title={Chatglm: A family of large language models from glm-130b to glm-4 all tools},
  author={Zeng, Aohan and Xu, Bin and Wang, Bowen and Zhang, Chenhui and Yin, Da and Zhang, Dan and Rojas, Diego and Feng, Guanyu and Zhao, Hanlin and others},
  journal={arXiv preprint arXiv:2406.12793},
  year={2024}
}

@inproceedings{ray,
  title={Ray: A distributed framework for emerging $\{$AI$\}$ applications},
  author={Moritz, Philipp and Nishihara, Robert and Wang, Stephanie and Tumanov, Alexey and Liaw, Richard and Liang, Eric and Elibol, Melih and Yang, Zongheng and Paul, William and Jordan, Michael I and others},
  booktitle={13th USENIX symposium on operating systems design and implementation (OSDI 18)},
  pages={561--577},
  year={2018}
}

@inproceedings{tong2025metamorph,
  title={Metamorph: Multimodal understanding and generation via instruction tuning},
  author={Tong, Shengbang and Fan, David and Li, Jiachen and Xiong, Yunyang and Chen, Xinlei and Sinha, Koustuv and Rabbat, Michael and LeCun, Yann and Xie, Saining and Liu, Zhuang},
  booktitle={Proceedings of the IEEE/CVF International Conference on Computer Vision},
  pages={17001--17012},
  year={2025}
}

@inproceedings{dong2023dreamllm,
  title={DreamLLM: Synergistic Multimodal Comprehension and Creation},
  author={Dong, Runpei and Han, Chunrui and Peng, Yuang and Qi, Zekun and Ge, Zheng and Yang, Jinrong and Zhao, Liang and Sun, Jianjian and Zhou, Hongyu and Wei, Haoran and others},
  booktitle={The Twelfth International Conference on Learning Representations},
  year={2023}
}

@article{ge2024seed,
  title={Seed-x: Multimodal models with unified multi-granularity comprehension and generation},
  author={Ge, Yuying and Zhao, Sijie and Zhu, Jinguo and Ge, Yixiao and Yi, Kun and Song, Lin and Li, Chen and Ding, Xiaohan and Shan, Ying},
  journal={arXiv preprint arXiv:2404.14396},
  year={2024}
}

@article{Step-audio,
  title={Step-audio: Unified understanding and generation in intelligent speech interaction},
  author={Huang, Ailin and Wu, Boyong and Wang, Bruce and Yan, Chao and Hu, Chen and Feng, Chengli and Tian, Fei and Shen, Feiyu and Li, Jingbei and Chen, Mingrui and others},
  journal={arXiv preprint arXiv:2502.11946},
  year={2025}
}

@inproceedings{panayotov2015librispeech,
  title={Librispeech: an asr corpus based on public domain audio books},
  author={Panayotov, Vassil and Chen, Guoguo and Povey, Daniel and Khudanpur, Sanjeev},
  booktitle={2015 IEEE international conference on acoustics, speech and signal processing (ICASSP)},
  pages={5206--5210},
  year={2015},
  organization={IEEE}
}

@article{van2017neural,
  title={Neural discrete representation learning},
  author={Van Den Oord, Aaron and Vinyals, Oriol and others},
  journal={Advances in neural information processing systems},
  volume={30},
  year={2017}
}

@article{geng2025x,
  title={X-omni: Reinforcement learning makes discrete autoregressive image generative models great again},
  author={Geng, Zigang and Wang, Yibing and Ma, Yeyao and Li, Chen and Rao, Yongming and Gu, Shuyang and Zhong, Zhao and Lu, Qinglin and Hu, Han and Zhang, Xiaosong and others},
  journal={arXiv preprint arXiv:2507.22058},
  year={2025}
}

@inproceedings{huang2024vbench,
  title={Vbench: Comprehensive benchmark suite for video generative models},
  author={Huang, Ziqi and He, Yinan and Yu, Jiashuo and Zhang, Fan and Si, Chenyang and Jiang, Yuming and Zhang, Yuanhan and Wu, Tianxing and Jin, Qingyang and Chanpaisit, Nattapol and others},
  booktitle={Proceedings of the IEEE/CVF Conference on Computer Vision and Pattern Recognition},
  pages={21807--21818},
  year={2024}
}

@article{soomro2012ucf101,
  title={Ucf101: A dataset of 101 human actions classes from videos in the wild},
  author={Soomro, Khurram and Zamir, Amir Roshan and Shah, Mubarak},
  journal={arXiv preprint arXiv:1212.0402},
  year={2012}
}

@inproceedings{bossard2014food,
  title={Food-101--mining discriminative components with random forests},
  author={Bossard, Lukas and Guillaumin, Matthieu and Van Gool, Luc},
  booktitle={European conference on computer vision},
  pages={446--461},
  year={2014},
  organization={Springer}
}

@inproceedings{huang2022prodiff,
  title={Prodiff: Progressive fast diffusion model for high-quality text-to-speech},
  author={Huang, Rongjie and Zhao, Zhou and Liu, Huadai and Liu, Jinglin and Cui, Chenye and Ren, Yi},
  booktitle={Proceedings of the 30th ACM International Conference on Multimedia},
  pages={2595--2605},
  year={2022}
}

@article{huang2025ddit,
  title={DDiT: Dynamic Resource Allocation for Diffusion Transformer Model Serving},
  author={Huang, Heyang and Hu, Cunchen and Zhu, Jiaqi and Gao, Ziyuan and Xu, Liangliang and Shan, Yizhou and Bao, Yungang and Ninghui, Sun and Zhang, Tianwei and Wang, Sa},
  journal={arXiv preprint arXiv:2506.13497},
  year={2025}
}
